\documentclass[a4paper,10pt]{article}

\usepackage[utf8]{inputenc}        
\usepackage{graphicx}              
\usepackage{url}                   
\usepackage{amsmath}               
\usepackage{cite}                  

\usepackage{geometry}              
\usepackage{titlesec}              
\usepackage{fancyhdr}              
\usepackage{setspace}              
\usepackage{helvet}                
\usepackage[table]{xcolor}         
\usepackage{etoolbox}              
\usepackage{parskip}               

\usepackage{soul}

\titleformat{\section}{\normalfont\bfseries\fontsize{11}{13}\selectfont}{\thesection.}{1em}{}
\titleformat{\subsection}{\normalfont\bfseries\fontsize{10}{12}\selectfont}{\thesubsection.}{1em}{}

\apptocmd{\thebibliography}{\setlength{\itemindent}{0.8cm}\setlength{\labelsep}{0.4cm}}{}{}

\begin{document}

{\centering
{\fontsize{14}{16}\selectfont\bfseries\MakeUppercase{In-Lab Carrier Aggregation Testbed for Satellite Communication Systems}} \\[1.5em]

\fontsize{11}{13}\selectfont
Jorge L. Gonz\'alez-Rios, Eva Lagunas, Hayder Al-Hraishawi, Luis M. Garc\'es-Socarr\'as, Symeon Chatzinotas \\
SnT - University of Luxembourg, Luxembourg\\
Corresponding Authors: jorge.gonzalez@uni.lu, eva.lagunas@uni.lu,luis.garces@uni.lu\\
}\vspace{1.5em}

\section{Abstract}
Carrier Aggregation (CA) is a technique used in 5G and previous cellular generations to temporarily increase the data rate of a specific user during peak demand periods or to reduce carrier congestion. 
CA is achieved by combining two or more carriers and providing a virtual, wider overall bandwidth to high-demand users of the system. 
CA was introduced in the 4G/LTE wireless era and has been proven effective in 5G as well, where it is said to play a significant role in efficient network capacity management. 
Given this success, the satellite communication (SatCom) community has put its attention into CA and the potential benefits it can bring in terms of better spectrum utilization and better meeting the user traffic demand. 
While the theoretical evaluation of CA for SatCom has already been presented in several works, this article presents the design and results obtained with an experimentation testbed based on Software Defined Radio (SDR) and a satellite channel emulator. 
We first present the detailed implementation design, which includes a Gateway (GW) module responsible for PDU-scheduling across the aggregated carriers, and a User Terminal (UT) module responsible for aggregating the multiple received streams. 
The second part of the article presents the experimental evaluation, including CA over a single Geostationary (GEO) satellite, CA over a single Medium Earth Orbit (MEO) satellite, and CA combining carriers sent over GEO and MEO satellites.
A key contribution of this work is the explicit consideration of multi-orbit scenarios in the testbed design and validation.
The testing results show promising benefits of CA over SatCom systems, motivating potential upcoming testing on over-the-air systems.

\section{Introduction}
\label{intro}

Multi-path transmission is a long-standing topic in wireless communication \cite{7501871}. 
Since its definition in 3GPP Release 10, carrier aggregation (CA) has become a key technology component in LTE and 5G cellular communication systems \cite{ca_lte}. 
Its popularity has rapidly increased in network deployments primarily because of its ability to allow operators to exploit fragmented licensed spectrum into marketable, higher data rates.
CA is the association of multiple carriers to/from a single user terminal and has become a key element employed in cellular wireless networks, enabling much higher data rates in response to sudden users' peak data demands. 
In the Satellite Communications (SatCom) domain, similar to the terrestrial one, providing a service tailored to the users' demands has been identified as a key functionality for modern systems. 
Adapting the offered capacity to the actual demand allows for efficient exploitation of the limited satellite resources while tailoring the offered capacity to different services with diverse and conflicting quality of service requests. 
In addition, SatCom systems typically face a non-uniform geographical distribution of the broadband traffic demand due to the wide coverage area \cite{7996648,Oltjon_2020}. 

It is important to keep in mind that CA does not add additional capacity to the system, unlike other PHY-based techniques, such as multi-beam/multi-satellite precoding \cite{ha2024usercentric, eappen2025livesat16, marrero2024sync, gonzalez2024cgd} and multi-satellite diversity combining \cite{singh2024diversity,CL2024time}. 
Compared to these advanced PHY techniques, CA offers lower implementation complexity and easier integration since it avoids the need for tight synchronization or complex joint processing. 
While precoding and diversity combining can improve capacity and robustness, they require accurate channel knowledge and higher processing overhead.
CA thus provides a practical and flexible way to enhance resource allocation without the operational challenges of those methods.
CA shall be seen as an additional degree of flexibility which can help in particular situations to better exploit and balance the satellite resource utilization (e.g., in the presence of congested and under-loaded channels). 
While the literature has plenty of works on long-term radio resource optimization for SatCom systems \cite{freedman,aravanis,cocco}, in this work, we focus on a short-term case, considering CA as a solution to user peak data demands. 

The application of the CA technique to SatCom systems has been investigated within the context of the ESA-funded activity CADSAT \cite{CADSAT_Project}, where the benefits and drawbacks of CA have been analyzed through the development of a CA Demonstrator laboratory testbed. This paper describes the CADSAT testbed in detail and presents a summary of the results achieved within the in-lab experimental campaign. Our previous works presented the general CA scenarios (with pros and cons) and an optimization framework for a multi-user multi-beam scenario \cite{kibria2,Kibria2020_short}. In both cases, results were limited to MATLAB-based system-level simulations.

The \textbf{main contributions} of this paper reside in the practical results obtained with the CADSAT in-Lab demo developed in \cite{CADSAT_Project}, based on Software Defined Radio (SDR) and a satellite channel emulator, and the explicit consideration of multi-orbit satellite scenarios in the system design and experimental validation. 
In particular, we first present the detailed implementation design, followed by the experimental evaluation including CA over a single Geostationary (GEO) satellite, CA over a single Medium Earth Orbit (MEO) satellite, and CA combining carriers sent through both GEO and MEO satellites.

The rest of this paper is organized as follows. Section II presents the satellite CA demonstrator as it was implemented in the SnT facilities at the University of Luxembourg. Section III explains the fundamental block of the testbed, which is the load balancing and Protocol Data Unit (PDU) scheduler block. Section IV presents a subset of the most relevant results achieved during the testing campaign. Section V concludes the paper.

\section{Satellite carrier aggregation demonstrator description}
\label{sec:testbed}

\begin{figure*}[htbp!]
\centering
\includegraphics[width=\textwidth]{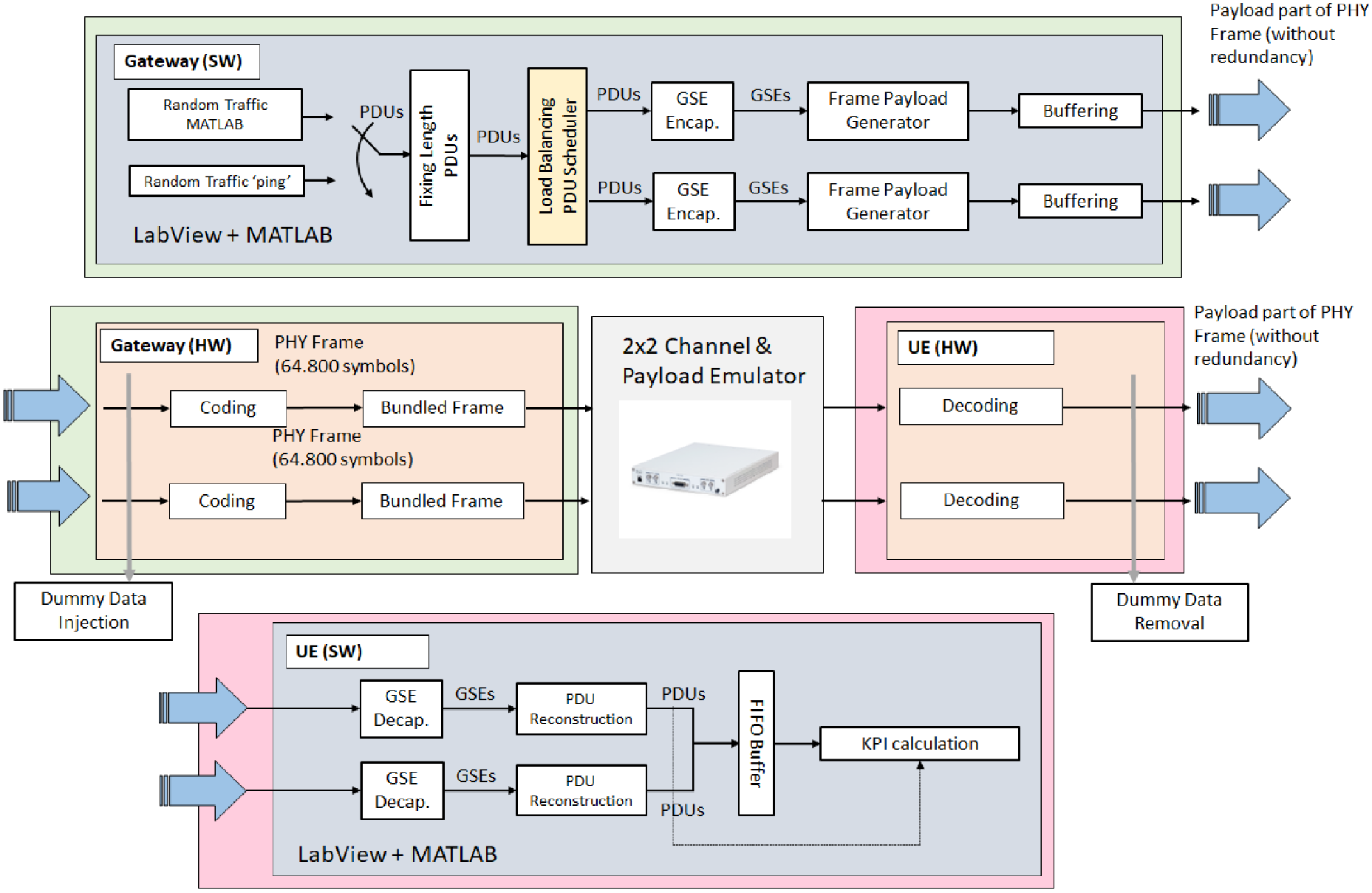}
\caption{Implementation scheme of the satellite carrier aggregation demonstrator.}
\label{fig:CADSAT_Demo}
\end{figure*}

Figure \ref{fig:CADSAT_Demo} depicts the implementation scheme of the satellite CA demonstrator developed in this work. 
The testbed implements CA for a single User Terminal (UT) over two carrier frequencies (which may be in different bands, have different bandwidths, and operate on different modulation and coding schemes).
In particular, Figure \ref{fig:CADSAT_Demo} shows a CA gateway (GW) (transmitter), a multi-orbit satellite channel and payload emulator, and a CA UT (receiver). 
Note that the receiver side has been intentionally kept as simple as possible with a traffic merging system consisting of a First In, First Out (FIFO) strategy. 
The reason is that satellite receivers shall be kept simple and low-cost, and the increased complexity can be accommodated in the GW side, thus reducing the overall cost of an eventually larger satellite network. 

At the GW side, a traffic generator produces the binary data (bits) to be transmitted, consisting of a series of fixed-length PDU.
The PDU goes through the ``Load Balancing and PDU Scheduling" block, which splits the traffic into two streams (two aggregated carriers)
, accounting for the characteristics of both links to ensure that proper packet ordering is perceived at the receiver side. 
Particularly, it considers the additional latency of the different multi-orbit links, i.e., the different propagation delays in the GEO and MEO paths.
Each carrier follows the conventional Generic Stream Encapsulation (GSE) and PHY framing procedures defined in the DVB-S2 standard \cite{DVBS2.2014}. 
The implementation of an adaptive buffer ensures a parallel transmission between the two links. 
Next, the traffic goes to the PHY layer, consisting of two Universal Software Radio Peripherals (USRPs), where each carrier is processed independently to form the corresponding Super Frame (SF) symbols and obtain the RF-modulated signals.
The resulting waveforms pass through the in-house developed channel emulator, which injects the noise and introduces the multi-orbit effects (differential/variable delays).
At the receiver, two USRPs synchronize, demodule, and decode the carriers (one each), after which the software runs the GSE decapsulation independently at each stream.
Finally, the PDUs are reconstructed and combined in a single FIFO buffer.

\subsection{SW-Part of the In-Lab Testbed}
\label{sec:testbedSW}
Upper and lower parts of Figure \ref{fig:CADSAT_Demo} illustrate the SW part of the testbed. 
There is a SW part in the gateway side emulating higher layers, i.e., traffic generation, scheduling of PDUs across different carriers, GSE encapsulation, payload part of the Base-Band Frame (BBFrame) preparation, and buffering. 
Buffering is very important as it takes care of two main tasks: (i) Make sure we have simultaneous transmission of data in both carriers; and (ii) Make sure that packets are pulled out from the PHY channel at the same time instance. 
It guarantees the synchronization between the two carriers, which is needed to avoid PDU ordering problems at the RX side.

\subsection{HW-Part of the In-Lab Testbed}
\label{sec:testbedHW}

\begin{figure}[htb]
\centering
\includegraphics[width=0.7\linewidth]{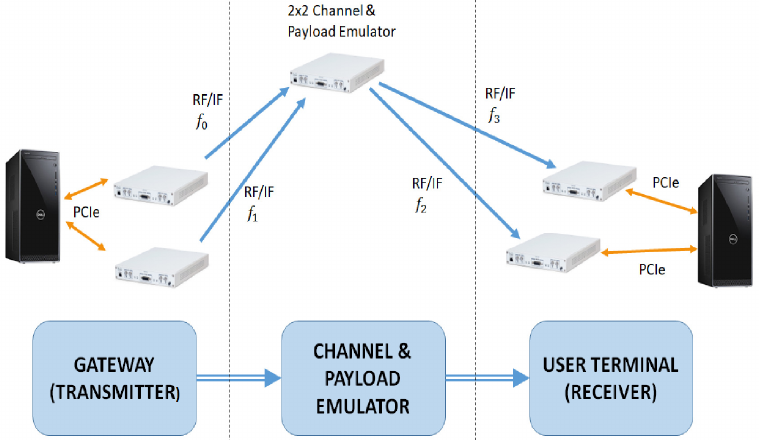}
\caption{Scheme of HW part}
\label{HWPart_fig}
\end{figure}

Figure \ref{HWPart_fig} illustrates the HW part of the testbed. 
Essentially, 2 x 2 Universal Software Radio Peripherals (USRPs) have been used to emulate the two pairs of carriers. Different USRPs are needed to emulate carriers of different bandwidths. 
Note that, while the synchronization between USRP-TX and USRP-RX is achieved with the conventional pilots of the Super-Frame (SF), there is no synchronization across the two carriers.
The functions of the HW part include:
\vspace{-9pt}
\begin{itemize}
  \item Take the BBFrames from the SW blocks and build the SF
  \vspace{-9pt}
  \item Modulation and Coding
  \vspace{-9pt}
  \item Channel emulation to achieve the desired SNR and multi-orbit effects 
  \vspace{-9pt}
  \item Synchronization of each carrier independently
  \vspace{-9pt}
  \item Receiving SFs and sending the BBFrames to the SW blocks at the RX side
  \vspace{-9pt}
  \item Sending dummy SF when there is no traffic available coming from the SW blocks
\end{itemize}

\section{Load Balancing and PDU Scheduler}
\label{sec:sche}

A key component in the GW part of Figure \ref{fig:CADSAT_Demo} is the load balancing and PDU scheduler, which needs to decide to which carrier to send the incoming PDUs. This decision will have a strong impact on the way the packets are received at the UT side, potentially causing packet disordering issues. The goal is to design a PDU scheduler that takes into account the characteristics of both links to ensure proper packet ordering at the receiver side. 
The inputs of the load balancing and PDU scheduler are:
\vspace{-9pt}
\begin{itemize}
  \item The input stream of PDUs (traffic).
  \vspace{-9pt}
  \item The signal-to-noise ratio (SNR) of the user in the 2 aggregated carriers, SNR$_1$, SNR$_2$. 
  \vspace{-9pt}
  \item The ``fill-rate'' of the users in the 2 aggregated carriers, FR$_1$, FR$_2$. The fill-rate essentially indicates the percentage of each carrier capacity that can be used by the user. For instance, setting FR$_1$=0.2 and FR$_2$=0.5 indicates that the UT is using 50\% of the capacity of carrier 2 and 20\% of the same in carrier 1. The fill-rate is an abstraction of the multiple UTs operating simultaneously with the UT under study.
\end{itemize}

 For the sake of simplicity, the load balancing scheduler does not consider fragmentation of PDUs across different BBFrames. This is a simplification that we opt for after considering the performance-complexity trade-off. It shall be remarked that the DVB standard does allow fragmentation of PDUs across different BBFrames.

We make use of the ``load balancing'' factor $\alpha$ to characterize the 2-carrier CA scenario. The definition of $\alpha$ is given as,
\begin{equation} 
 \alpha = \frac{C_2 \cdot FR_2}{C_1 \cdot FR_1} = \frac{BW_2\cdot (M_2 \cdot CR_2) \cdot FR_2}{BW_1\cdot (M_1 \cdot CR_1) \cdot FR_1} 
 \label{alpha_eq}
\end{equation}
where the key parameters are summarized in Table \ref{param_tab}. Note that it is assumed that carrier 1 is the dominant carrier with $C_1 \geq C_2$, and $FR_1 \neq 0$.

\begin{table}[htb]
\centering
\caption{Key Parameters}
\label{param_tab}
  \begin{tabular}{| c | c | }
	\hline
\textbf{Parameter} & \textbf{Description} \\
	\hline
$\alpha$ &    Load Balancing Factor (between 0 and 1) \\	   \hline
$C_1$    &    Capacity of carrier 1 [bps]   \\	   \hline
$C_2$    &    Capacity of carrier 2 [bps]  \\	   \hline
$FR_1$   &    Fill Rate of Carrier 1 (between 0 and 1)  \\	   \hline
$FR_2$   &    Fill Rate of Carrier 2 (between 0 and 1)  \\	   \hline
$BW_1$   &    Bandwidth of Carrier 1 [Hz]  \\	   \hline
$BW_2$   &    Bandwidth of Carrier 2 [Hz]  \\	   \hline
$M_1$    &    Modulation Order of carrier 1  \\	   \hline
$M_2$    &    Modulation Order of carrier 2 \\	   \hline
$CR_1$   &    Code Rate  \\	   \hline
$CR_2$   &    Code Rate   \\
	\hline
  \end{tabular}
\end{table}

For the sake of fast execution and after investigating multiple alternatives, the design of the PDU scheduling was implemented based on a limited set of $\alpha$ recorded in a look-up table, which is illustrated in Table \ref{Scheduling_Seq_fig} for completeness. Table \ref{Scheduling_Seq_fig} shows load balancing factor $\alpha$ in column-left, while column-right indicates the scheduling sequence, where ``1'' means PDU to carrier 1, and ``2'' means PDU to carrier 2. It should be noted that a change in modulation and coding sequence will result in a different modulation order and code rate, thereby claiming a change in the scheduling sequence. Since all is managed at the GW side, no major problems are expected in this regard.

\begin{table}\caption{Look up table: PDU scheduling sequence based on load-balancing factor} \label{Scheduling_Seq_fig}   
\centering

\begin{tabular}{|>{\centering\arraybackslash}m{0.19\linewidth} |m{0.72\linewidth}|}
\hline
\textbf{\cellcolor{blue!25}Balanced Factor $\alpha$} & \textbf{\cellcolor{blue!25}Scheduling sequence} \\ \hline

0.2  & [1,1,1,1,1,2] \\ \hline 
0.25 & [1,1,1,1,2] \\ \hline 
0.3  & [1,1,1,2,1,1,1,2,1,1,1,1,2] \\ \hline 
0.35 & [1,1,2,1,1,1,2,1,1,1,2,1,1,1,2,1,1,1,2,1,1,1,2,1,1,1,2] \\ \hline 
0.4  & [1,1,2,1,1,1,2] \\ \hline 
0.45 & [1,1,2,1,1,2,1,1,2,1,1,2,1,1,1,2,1,1,2,1,1,2,1,1,2,1,1,1,2] \\ \hline 
0.5  & [1,1,2] \\ \hline 
0.55 & [1,2,1,1,2,1,1,2,1,1,2,1,1,2,1,2,1,1,2,1,1,2,1,1,2,1,1,2,1,1,2] \\ \hline 
0.6  & [1,2,1,1,2,1,1,2] \\ \hline 
0.65 & [1,2,1,1,2,1,2,1,2,1,2,1,1,2,1,2,1,1,2,1,2,1,1,2,1,2,1,1,2,1,1,2] \\ \hline 
0.7  & [1,2,1,2,1,1,2,1,2,1,1,2,1,2,1,1,2] \\ \hline 
0.75 & [1,2,1,2,1,1,2] \\ \hline 
0.8  & [1,2,1,2,1,2,1,1,2] \\ \hline 
0.85 & [1,2,1,2,1,2,1,2,1,2,1,1,2,1,2,1,2,1,2,1,2,1,2,1,1,2,1,2,1,2,1,2,1,2,1,1,2] \\ \hline 
0.9  & [1,2,1,2,1,2,1,2,1,2,1,2,1,2,1,2,1,1,2] \\ \hline 
0.95 & [1,2,1,2,1,2,1,2,1,2,1,2,1,2,1,2,1,2,1,2,1,2,1,2,1,2,1,2,1,2,1,2,1,2,1,2,1,1,2] \\ \hline 
1    & [1,2] \\ \hline 

\end{tabular}

\end{table}

\subsection{Load Balancing Scheduling for Multi-Orbit Scenario:}
\label{par:sche2}

The load balancing scheduling should consider the additional latency of the different multi-orbit links, e.g., the GEO link with respect to the MEO link. In other words, the load-balancing scheduler needs to take into account the differential propagation delay. We consider that the GW starts transmitting over both MEO and GEO links at the same time, and in a way that packets are received in the appropriate order at the UT side.
For illustration purposes, let us assume that the differential propagation delay corresponds to 5 PDUs and that both MEO and GEO links have the same characteristics in terms of SNR and bandwidth. 
Therefore, the first 5 PDUs shall be sent to the MEO carrier, and after that, we shall assign 1 PDU to each carrier. 

Considering an average Gateway-Satellite-UT trip time of 2 $\times$ 11.933 km for the MEO carrier and 2 $\times$ 40.151 km for the GEO carrier, this results in a differential propagation delay between the GEO and the MEO satellite of 2 $\times$ 28.218 km, or 188.1 ms. For a carrier bandwidth $BW$, the number of superframes (SFs) transmitted during this time can be calculated as,
\begin{equation}
 N_{SF}=\frac{188.1~\text{[ms]}}{T_{SF}~\text{[ms]}}
\end{equation}
where $T_{SF} = \frac{612540~\text{symbols}}{BW}$ is the SF duration. For instance, assuming $BW = 5$ MHz (4.64 MHz in practice due to roll-off factor), we obtain $T_{SF} =  0.1225$ sec, and $N_{SF} = 1.425$ SFs.

Accounting for 9 bundled frames in each SF with format 2 (as defined in the standard \cite{DVBS2X_fullStd}) and a constant MODCOD with modulation order $2^M$, the amount of transmitted FECFRAMEs in the interval is $9\times M\times N_{SF}$.
Then, given how many PDUs carry the FECFRAME $N_\text{PDU/FECFRAME}$ (which depends on the PDU size and carrier fill rate), the scheduler must assign the following number of PDUs to the MEO carrier at the beginning of the transmission: 
\begin{equation*}
\text{Initial sequence length for MEO carrier} =  \\ 
N_\text{PDU/FECFRAME}\times 9\times M\times \frac{BW_\text{MEO} \times 188.1\times10^{-3} }{612540~\text{symbols}}
\end{equation*}

For example, consider $SNR=10$ dB, which corresponds to ModCod 8PSK 5/6, i.e., $M=3$ bits/symbol, and considering $BW=4.64$ MHz) and $FR=0.25$, we have $N_\text{PDU/FECFRAME}=1$. Therefore:
\begin{equation*}
   \text{Initial sequence length for MEO carrier} = \\ 1 \times 9\times 3 \times \frac{4.64 \times10^{6} \times 188.1\times10^{-3} }{612540~\text{symbols}} = 38.4712.
\end{equation*}

In such a case, the first 38 packets shall be put on the MEO carrier queue. 
Although the MEO propagation path varies, its variation occurs at a slow pace due to the high-altitude orbit (approx. 8000 km). Therefore, we take an average propagation delay as the trip time to obtain the initial sequence length.

\section{Testbed Results}
\label{sec:simu}

\begin{figure}[h!]
\begin{minipage}[c]{0.49\linewidth}
  \centering
  \centerline{\includegraphics[scale=0.55]{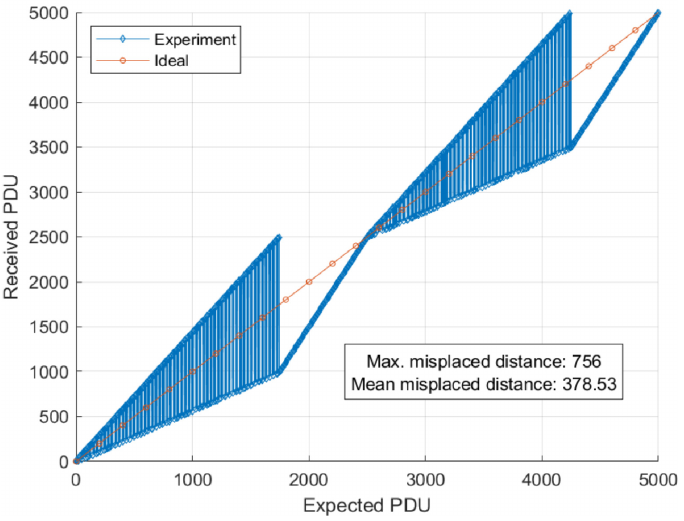}}
  \centerline{\small(a) Benchmark Round Robin (RR)} 
\end{minipage}
\begin{minipage}[c]{0.49\linewidth}
  \centering
  \centerline{\includegraphics[scale=0.55]{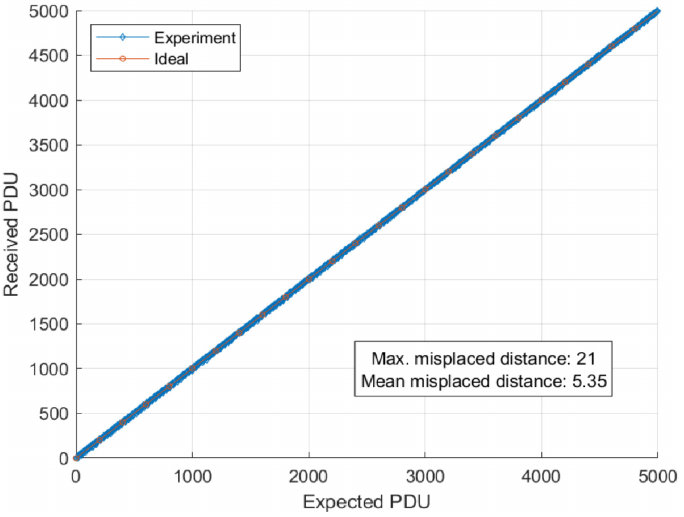}}
  \centerline{\small(b) Proposed Load Balancing (LB) scheduling} 
\end{minipage}
\caption[Text excluding the matrix]{Results regarding ordering of PDUs: Carrier 1 with  $BW_1 = 5$ MHz and carrier 2 with  $BW_1 = 2$ MHz. Load balancing factor $\alpha$ equal to $0.4$. }
\label{Test1.5_fig}
\end{figure}

As first results, we show the performance of CA with Round Robin (RR) scheduling and with the proposed Load Balancing (LB) scheduling (Section \ref{sec:sche}) for a single user performing CA over two carriers, whose only difference is in the bandwidth. 
In particular, we investigate the case of CA in two carriers with the same SNR (i.e. $\text{SNR}_1 = \text{SNR}_2 = 10$ dB), and same fill-rate (i.e. $FR_1 = FR_2 = 0.25$) while using different bandwidth (i.e. $BW_1 \geq BW_2$). More specifically, we have carrier 1 with  $BW_1 = 5$ MHz and carrier 2 with  $BW_1 = 2$ MHz. 
As a consequence, the load balancing factor $\alpha$ is equal to $0.4$. 
Figure \ref{Test1.5_fig} shows the relationship between received PDUs and their expected order for both RR and LB scheduling.

It can be quickly observed that RR is experiencing problems with the PDU ordering, as the experimental curve (in blue) moves away from the ideal slope curve (in red). On the other hand, LB scheduling behaves correctly, with some slight misplacement errors due to the unbalanced situation. 
Note that from Figure \ref{Test1.5_fig}(a), the 2 phases of the experiment can be clearly observed: The total 5.000 PDUs have been transmitted in 2 bunches of 2.500 PDUs each (for implementation purposes). 
It can be seen that the errors accumulate over time. 
This is evident from the blue cone that appears towards the end of each burst transmission. If we let the experiment run longer, we would definitely observe more errors. 
The numbers indicate that those PDUs that are received in incorrect order appear to be 378 positions away (on average) from their expected location when using RR. 
In contrast, the number reduces to close to 5 when considering the proposed LB scheduling.
Figure \ref{Test1.5_fig} evidences a general trend observed during the experimental campaign: (i) Round-robin is not able to adapt to the imbalance of the carriers, thus producing ordering errors at the receiver; and (ii) the proposed scheduling can provide close to ideal behavior. 
Table \ref{test1_geoca_misplacement_tab} compares other load balancing factors.

\begin{table}[h!]
\centering
\caption{CADSAT Test 1 - GEO CA: PDU misplacement errors}
\label{test1_geoca_misplacement_tab}
\resizebox{.65\columnwidth}{!}{%
\begin{tabular}{| c  | c | c | c | c | c | c | c |  }
	\hline
\cellcolor{blue!25}  &  \cellcolor{blue!25}  & \cellcolor{blue!25}  $\alpha=1$ & \cellcolor{blue!25} $\alpha=0.9$ & \cellcolor{blue!25} $\alpha=0.7$ & \cellcolor{blue!25} $\alpha=0.6$ & \cellcolor{blue!25} $\alpha=0.4$ & \cellcolor{blue!25} $\alpha=0.3$ \\
	\hline
LB   &  Mean   & 0.76	& 3.72   & 1.96   &  2.15   & 5.35  &  3.09   	 \\	  
  \cline{2-8} &  Max    &  3  	&   10   &   16   &    19   &   21	&   20	      \\	  
	\hline
RR   &  Mean   & 0.76	& 64.01   & 189.16   &  251.14   & 378.53  &  439.86   	 \\	  
  \cline{2-8} &  Max    &  3  	&   128   &   379   &    501   &   756	&   879	      \\	 
	\hline
  \end{tabular}
  }
\end{table}

\begin{figure*}[htbp!]
\centering
\includegraphics[width=\textwidth]{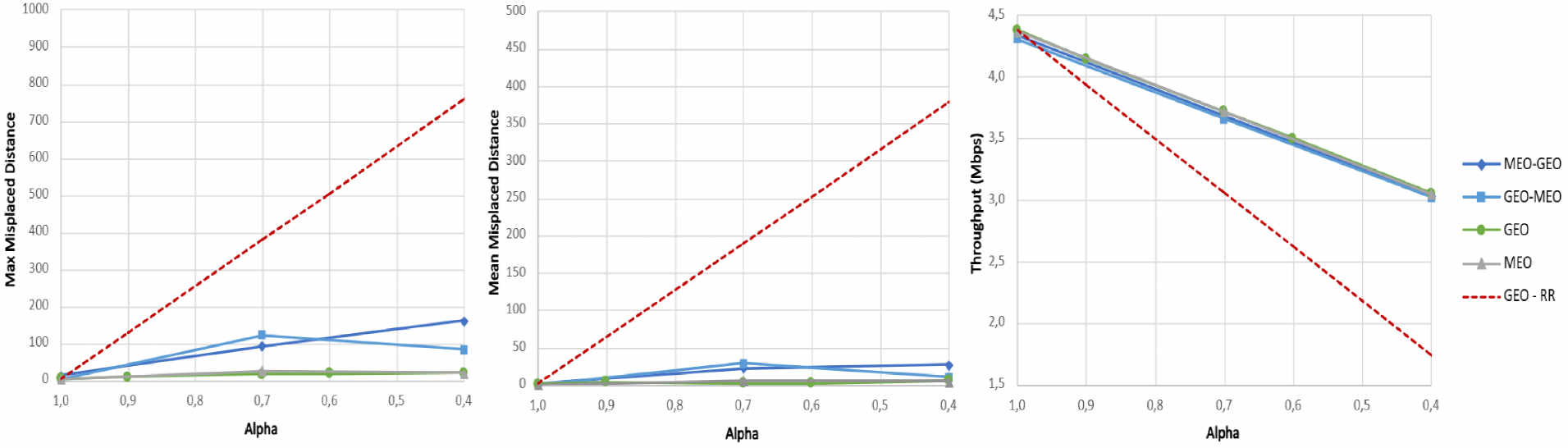}
\caption{Overall comparison between CA scenarios: (Left) Maximum misplaced packet distance; (Center) Mean misplaced packet distance; and (Right) Achievable throughput in Mbps.}
\label{final_fig}
\end{figure*}

The experimental campaign considered a comparison among the different scenarios, listed below for completeness:\vspace{-6pt}
\begin{itemize}
  \item GEO CA with the proposed scheduling (indicated herein as GEO)
  \vspace{-9pt}
  \item GEO CA with round robin scheduling (indicated herein as GEO-RR)
  \vspace{-9pt}
  \item MEO CA with the proposed scheduling (indicated herein as MEO)
  \vspace{-9pt}
  \item MEO-GEO CA with the proposed scheduling where MEO carrier is faster than the GEO carrier (indicated herein as MEO-GEO). We set same SNR and FR (i.e. $\text{SNR}_1 = \text{SNR}_2 = 10$ dB, $FR_1 = FR_2 = 0.25$), we fixed $BW_1$ for MEO case to 5 MHz and play with the value of $BW_2$ for the GEO case which is $BW_2 \leq BW_1$.
  \vspace{-9pt}
  \item GEO-MEO CA with the proposed scheduling where GEO carrier is faster than the MEO carrier (indicated herein as GEO-MEO). Parameters are same as before but assuming that GEO is carrier 1, and MEO is carrier 2.
\end{itemize}
It shall be noted that, while the load balancing design considers a fixed average trip-time delay for the MEO orbit, the testbed emulates the actual varying MEO trip-time.
Figure \ref{final_fig} presents the overall comparison in terms of max. misplace distance (ordering error distance in number of PDU packets), mean misplace distance and achievable aggregated throughput. We can observe that the best results are achieved with the single-orbit CA (i.e. either MEO or GEO CA) with the proposed scheduling. In fact, the round robin scheduling is not able to adapt to the unbalanced carriers (i.e. $\alpha < 1$) and shows the worst performance. In between, we have the MEO-GEO and GEO-MEO CA, which show satisfactory performance although suffer a bit more from the multi-orbit latency compensation/adaptation. Finally, in terms of achievable throughput, all tested CA scenarios achieve similar results, with the single-orbit case providing a slightly higher throughput.

\section{Conclusions and Remarks}
\label{sec:conclusion}
This paper analyzes the benefits of the CA technique applied to satellite communications systems. In particular, this work presents the SDR-based testbed built during the ESA CADSAT project and reports the relevant results achieved during the testing campaign.
The learned take-home lessons are:
\vspace{-6pt}
\begin{itemize}
  \item The CA with LB scheduling behaves significantly better than RR, especially when the carriers are unbalanced.
  \vspace{-9pt}
  \item Although small errors appear when $\alpha < 1$ for the proposed load balancing scheduling, the TCP protocol might be able to account for such small reordering procedure.
  \vspace{-9pt}
  \item To obtain good results, the proposed look-up table with the scheduling sequence (presented in Table \ref{Scheduling_Seq_fig}) needs to accurately reflect the load balancing ratio between the two aggregated carriers.
  \vspace{-9pt}
   \item Regarding the multi-orbit CA, we have shown that it is necessary to add a prefix to the PDU scheduling sequence which corresponds to the number of PDUs that can be transmitted within the differential propagation delay of the two carriers. Based on the testing done in this activity, a single prefix works fine for carriers with BW around 5-7 MHz and for MEO-GEO CA. However, this depends on the BW of the carriers and the actual NGSO orbit. 
\end{itemize}

The results experimentally validate the feasibility of adopting CA in SatCom and pave the way for future over-the-air testing.
Future work will also consider a more compact modem design, potentially leveraging RF System-of-Chip (RFSoC) advantages for integrating the software and hardware blocks.

\section*{Acknowledgment}
This work has been supported by the European Space Agency (ESA) funded activity CADSAT: Carrier Aggregation in Satellite Communication Networks (4000122090/17/UK/ND). 
The views of the authors of this paper do not necessarily reflect the views of ESA.
The authors would like to thank J. Grotz from SES, Luxembourg, for their helpful advice on various practical aspects related to the application of CA on satellite systems. The authors would also like to thank the rest of Univ. of Luxembourg's team that participated in the CADSAT project: Rakesh Palisetty, Jevgenij Krivochiza, Jorge Querol, Juan C. Merlano Duncan and Sumit Kumar. The authors thank Nicolo Mazzali and Nikos Toptsidis, technical officers from ESA supervising the CADSAT activity. 

\section{References}
\vspace{-2em}
\bibliographystyle{naturemag}
\begingroup
\renewcommand
\refname{}
\setlength{\parskip}{0pt}
\bibliography{references}
\endgroup

\end{document}